\documentclass[twocolumn,pra]{revtex4-1}

 \usepackage{amssymb}
 \usepackage{dsfont}
 \usepackage{amsmath}
\usepackage{graphicx}% Include figure files
\usepackage{dcolumn}% Align table columns on decimal point
\usepackage{bm}% bold math

\begin{document}

\title{Exact Entanglement Dynamics Beyond the Rotating Wave Approximation }%\thanks{A footnote to the article title}%
\author{S. M. Hashemi Rafsanjani, S. Agarwal and J. H. Eberly}
\affiliation{ Rochester Theory Center and the Department of Physics
\& Astronomy\\
University of Rochester, Rochester, New York 14627}
\email{hashemi@pas.rochester.edu}
%\author{.........}
%\email{......@pas.rochester.edu}
%\author{ll}
%\author{}

%\preprint{}

\date{\today}

\begin{abstract} The entanglement dynamics of two remote qubits is examined analytically. The qubits interact arbitrarily strongly with separate harmonic oscillators in the idealized degenerate limit of the Jaynes-Cummings Hamiltonian. In contrast to well known non-degenerate RWA results, it is shown that ideally degenerate qubits cannot induce bipartite entanglement between their partner oscillators.
\end{abstract}

\pacs{..........}

% PACS
\maketitle


 \newcommand{\beq}{\begin{equation}}
 \newcommand{\eeq}{\end{equation}}
 \newcommand{\bel}{\begin{align*}}
 \newcommand{\tamam}{\end{align*}}
 \newcommand{\dg}[1]{#1^{\dagger}}
 \newcommand{\reci}[1]{\frac{1}{#1}}
 \newcommand{\ket}[1]{|#1\rangle}
 \newcommand{\nim}{\frac{1}{2}}
 \newcommand{\om}{\omega}
 \newcommand{\te}{\theta}
 \newcommand{\la}{\lambda}
 \newcommand{\beqa}{\begin{eqnarray}}             %Begin Equation Array
 \newcommand{\eeqa}{\end{eqnarray}}               %End Equation Array
 \newcommand{\nn}{\nonumber}                      %No Number
 \newcommand{\bra}[1]{\langle#1\vert}                 % Bra
 \newcommand{\ipr}[2]{\left\langle#1|#2\right\rangle}
  \newcommand{\up}{\uparrow}
   \newcommand{\down}{\downarrow}
     \newcommand{\dn}{\downarrow}         % Inner Product

% Introduction
\section{Introduction}

Entanglement is considered a necessary resource for many of the algorithms proposed for quantum computation and communication \cite{nielsen:2000}.
Over the course of the last decade there has been a growing interest in finding ways to quantify \cite{RevModPhys.81.865}, manipulate, and control 
\cite{RevModPhys.73.565} the initial entanglement shared by different parties when they come in contact with different  local \cite*{PhysRevLett.91.070402,*PhysRevLett.89.277901,*PhysRevA.65.040101,*PhysRevA.68.062316,*springerlink:10.1007/s11128-009-0137-6,*springerlink:10.1134/S0030400X10030069} and non-local \cite{PhysRevLett.93.140404,yonac1,PhysRevLett.99.160502,PhysRevA.77.012117,PhysRevLett.104.070406,PhysRevA.76.022312} environments. For two remote systems coming in contact with two uncorrelated reservoirs, typically the initial entanglement between the two systems ultimately ends up as a bipartite entanglement between the two reservoirs \cite{PhysRevLett.101.080503}. For the single mode environments, however, the entanglement dynamics depends strongly on the initial state of the two environments and the interaction between each system and the corresponding environment \cite{yonac1,yonac2010}.

In many of the previous investigations the Jaynes-Cummings (JC) model \cite{jcpaper}  has been invoked to describe the interaction between each party, described as qubits, and the corresponding environment, modeled by a harmonic oscillator. The JC Hamiltonian reads
\beq
H=\hbar\frac{\omega_0}{2}\sigma_z+\hbar\omega \dg{a} a +\hbar \la (a^{\dagger}+a)\sigma_x, \label{salam1}
\eeq
where $\sigma_{z}$ and $\sigma_{x}$ are Pauli matrices and $a$ and $a^{\dagger}$ are the usual ladder operators.
The model has been used extensively to describe the interaction between an atom and a single mode of a cavity in quantum optics \cite{jcpaper,eberly1987}. In the studies of strong light-matter interaction and/or in the search for potential quantum computation and quantum information applications, the model has also been invoked to describe the interaction between  a Cooper pair box with a nanomechanical resonator \cite{PhysRevB.68.155311} or with a transmission line resonator \cite{PhysRevA.69.062320,ISI:000223746000038}, etc. In quantum optics, typically, the nearly resonant ($|\omega-\omega_0|\ll\omega +\omega_0$) and weak coupling ($\la\ll\omega,\omega_0) $ conditions apply and the rotating wave approximation (RWA) is valid \cite{jcpaper,eberly1987}.  Yet with the advent of circuit QED it has become feasible experimentally to explore regimes of the model where the dynamics is not well described within the RWA \cite{Nature.6.772,PhysRevLett.105.060503,PhysRevLett.105.237001}.

There have already been many investigations exploring analytically and/or numerically the local dynamics of the model beyond the RWA  \cite{*[{}] [{ and references therein}] BWShorebook,0301-0015-6-2-010,*PhysRevA.46.4138,*PhysRevA.68.063811,*1402-4896-76-2-007,PhysRevA.37.1628,PhysRevA.50.2040,irishprl,irish2005,casanova,PhysRevA.82.062320} . Some of
the developed techniques deal with nearly resonant but strong couplings \cite{irishprl,PhysRevA.50.2040,PhysRevA.37.1628} and some deal with highly detuned and/or strong coupling scenarios \cite{irish2005,casanova,PhysRevA.82.062320}. 
Remote entanglement dynamics beyond the RWA has also been the subject of a recent note by Chen et al. \cite{chen}. They focused on the nearly resonant
and strong coupling scenario. In this note we present exact analytic formulas for the entanglement dynamics, beyond the RWA, in the far from resonance regime of the model where the qubits are degenerate $(\om_{0}=0)$. A study of the dynamics of a degenerate qubit interacting with a classical field, and a discussion of a physical system which can be treated as a degenerate qubit, has been given by Shakov and McGuire \cite{PhysRevA.67.033405}. 

\begin{figure}[b]
\includegraphics[width=8cm]{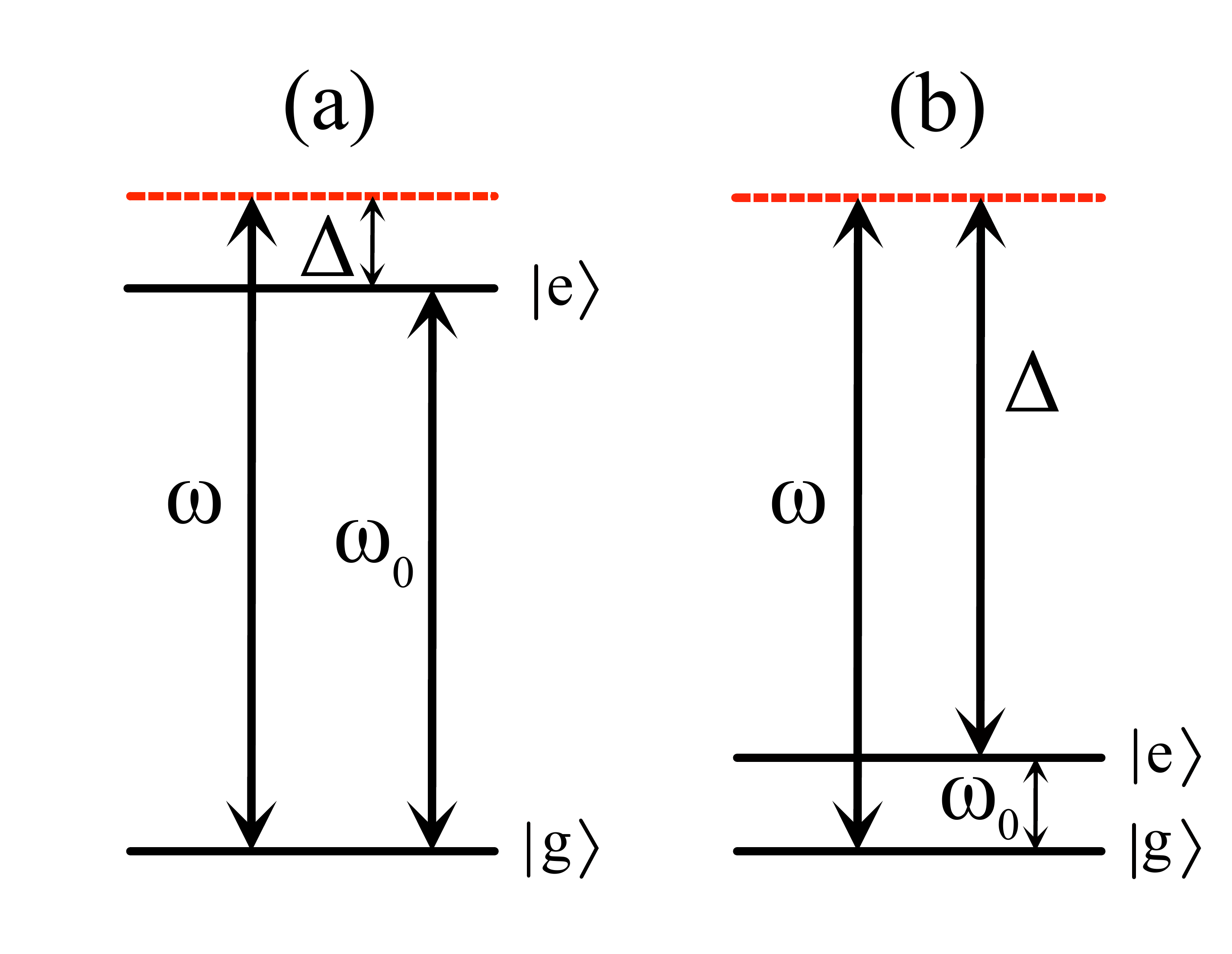}
\vspace{-.5cm}
\caption{The energy level representation of (a) RWA and (b) nearly degenerate regime. In the degenerate regime $\om_{0}=0$.}
\label{graph1}
\end{figure}

In Fig. \ref{graph1} we present a schematic comparison of the energy levels of the qubit and the harmonic oscillator in  both RWA and near-degenerate regimes. The system we study consists of two non-communicating subsystems. Each subsystem itself consists of a qubit that interacts with a partner harmonic oscillator. From now on we refer to the harmonic oscillators as the fields.

The structure of the current note is as follows. In section \ref{sec2} we briefly examine the JC Hamiltonian in the degenerate regime. Section \ref{sec3} is devoted
to the dynamics of a single qubit and a single harmonic oscillator in the degenerate regime where we completely avoid the RWA. In section \ref{sec4} we focus on a certain class of initial states to study the 
entanglement dynamics of two remote initially entangled qubits in the degenerate regime. The effect of different initial fields as well as the effect of coupling strength on the bipartite entanglement
in the degenerate regime is studied. Intuitively we may think that when two separable fields come in contact with remote entangled qubits, some of the coherence between the two qubits gets transferred to the fields and they develop bipartite entanglement \cite{yonac2007}. We show that in the degenerate regime two initially separable fields do not develop bipartite entanglement.

\section{Degenerate Regime}\label{sec2}
In the degenerate regime  $(\om_{0}=0)$, the JC Hamiltonian reads 
\beqa\nn
H&=\left(\hbar\om\dg{a}a+\hbar\la(\dg{a}+a)+\hbar\frac{\la^{2}}{\om} \right)\ket{\up}\bra{\up}\\&+\left(\hbar\om\dg{a}a-\hbar\la(\dg{a}+a) +\hbar\frac{\la^{2}}{\om} \right)\ket{\down}\bra{\down}. \label{salam2}
\eeqa
In this note $\sigma_{x}\ket{\up}=\ket{\up}$, $\sigma_{x}\ket{\dn}=-\ket{\dn}$, $\sigma_{z}\ket{e}=\ket{e}$ and $\sigma_{z}\ket{g}=-\ket{g}$. For simplicity 
we added a constant value ($\hbar\frac{\la^{2}}{\om}$) to the Hamiltonian that has no effect on the dynamics. $H$ can be diagonalized as follows:
\begin{align}\nn
&H \ket{E_{n,\pm}}=E_n\ket{E_{n,\pm}}
,~~~~~~~~E_n=n \hbar\om,\\  \nn
&\ket{E_{n,+}}=\ket{\up,n_{+}},\\  \nn
& \ket{n_{+}}=\dg{D}(\beta)\ket{n}=\dg{D}(\beta)\frac{{\dg{a}}^n}{\sqrt{n!}}\ket{0},\\  \nn
&\ket{E_{n,-}}=\ket{\dn,n_{-}},\\ \
&\ket{n_{-}}=\dg{D}(-\beta)\ket{n}=\dg{D}(-\beta)\frac{{\dg{a}}^n}{\sqrt{n!}}\ket{0}.
\end{align}
Here $\beta=\frac{\la}{\om}$, and $D(\beta)=\exp(\beta\dg{a}-\beta^{*}a)$ is the harmonic oscillator displacement operator. The set of all eigenstates provides a complete basis for the Hilbert space and the closure relation reads
\begin{align} 
\sum_{n}\Big( \ket{\up,n_{+}}\bra{\up,n_{+}}+\ket{\dn,n_{-}}\bra{\dn,n_{-}}\Big)=\openone.
\end{align}

%DYNAMICS OF DEGENERATE REGIME%
\section{Dynamics in degenerate regime}\label{sec3}
Here we focus on the dynamics of a single degenerate qubit interacting with a single harmonic oscillator. This is to emphasize that some of the features of the entanglement dynamics are generic consequences of the degenerate regime and not the 
specific setup we focus on next. The qubit and the field are assumed to be initially separable i.e. $\rho_{QF}(0)=Q\otimes F$  where  $Q$ and $F$ denote the qubit and field states respectively. Here we focus on the reduced density matrix 
of the qubit alone, $Q(t)=Tr_{F}[\rho_{QF}(t)]$, and ask how its matrix elements evolve in time. With a generic initial atomic state, the qubit density matrix can be written as 
\begin{align}\nn
\rho_{QF}(0)=(Q_{\up \up}\ket{\up}\bra{\up}+Q_{\dn \dn}\ket{\dn}\bra{\dn}+\\ Q_{\up \dn}\ket{\up}\bra{\dn}+Q_{\dn \up}\ket{\dn}\bra{\up})\otimes F. 
\label{salam3} 
\end{align}

As mentioned in the previous section $[\sigma_{x},H]=0$ and thus one can conclude that $Q_{\up\up}(t)=\bra{\up}Q(t)\ket{\up}=Q_{\up\up}$ and $Q_{\dn\dn}(t)=\bra{\dn}Q(t)\ket{\dn}=Q_{\dn\dn}$ and only $Q_{\up\dn}(t)$ and $Q_{\dn\up}(t)$ have nontrivial dynamics. Furthermore we
know that $Q_{\up\dn}(t)=Q_{\dn\up}(t)^{*}$.~The only contributing term to $Q_{\up\dn}(t)=\bra{\up}Q(t)\ket{\dn}$ comes from the propagation of the third term in Eq. (\ref{salam3}):\\
\begin{align}
Q_{\up\dn}(t)=Q_{\up\dn}\bra{\up}Tr_{F}\{U\ket{\up}\bra{\dn}\otimes F~ \dg{U}\}\ket{\dn},
\end{align}
where $U=e^{-iHt/\hbar}$ is the propagator. In Appendix A we have worked out the dynamics when the field was initially in a coherent state and the qubit is initially in either $\ket{\up}$ or $\ket{\dn}$. Thus, here we can invoke a diagonal coherent state representation of the field and the result of Appendix A to find $Q_{\up \dn}(t)$. By employing the Glauber-Sudarshan representation \cite{PhysRev.131.2766,PhysRevLett.10.277}: $$F=\int \text{d}^{2}\alpha P(\alpha)\ket{\alpha}\bra{\alpha}$$ one can rewrite  $Q_{\up\dn}(t)$ as the following: 
\begin{align}
Q_{\up\dn}(t)=Q_{\up\dn}\int\text{d}^{2}\alpha P(\alpha)\bra{\up}Tr_{F}\{U\ket{\up,\alpha}\bra{\dn,\alpha}\dg{U}\}\ket{\dn}.\label{salam4}
\end{align}\\
In Appendix A it is shown that: 
\begin{align}\nn
&U\ket{\up,\alpha}=\ket{\up,(\alpha+\beta)e^{-i \om t}-\beta} e^{-i\beta^{2}\sin\om t}e^{i\beta Im[\alpha\gamma^{*}(t)]}\\ 
&U\ket{\dn,\alpha}=\ket{\dn,(\alpha-\beta)e^{-i \om t}+\beta} e^{-i\beta^{2}\sin\om t}e^{i\beta Im[\alpha^{*}\gamma(t)]}\label{salam5}
\end{align}
where \begin{align}\gamma(t)=e^{i\om t}-1.\end{align}\\ The factor $\gamma(t)$ is a complex number that follows a circle around $-1$ in the complex plane. The field components of the above evolved states are coherent states. In deriving these relations we assumed $\beta$ to be real. Using identities in Eq. (\ref{salam5}) one can show that
\begin{align}\nn
Q_{\up\dn}(t)&=Q_{\up\dn}\int\text{d}^{2}\alpha P(\alpha) e^{\beta \alpha \gamma^{*}(t)}e^{-\beta \alpha^{*}\gamma(t)} \\  &~~~~~~~\ipr{(\alpha-\beta)e^{-i\om t}+\beta}{(\alpha+\beta)e^{-i \om t}-\beta} \label{salam6} \\
&=Q_{\up\dn}e^{-2\beta^{2}|\gamma(t)|^{2}}\int \text{d}^{2}\alpha P(\alpha)e^{4i\beta Im[\alpha\gamma^{*}(t)]}.
\label{salam7}
\end{align}
 \begin{figure}[htpb]
\includegraphics[width=\columnwidth]{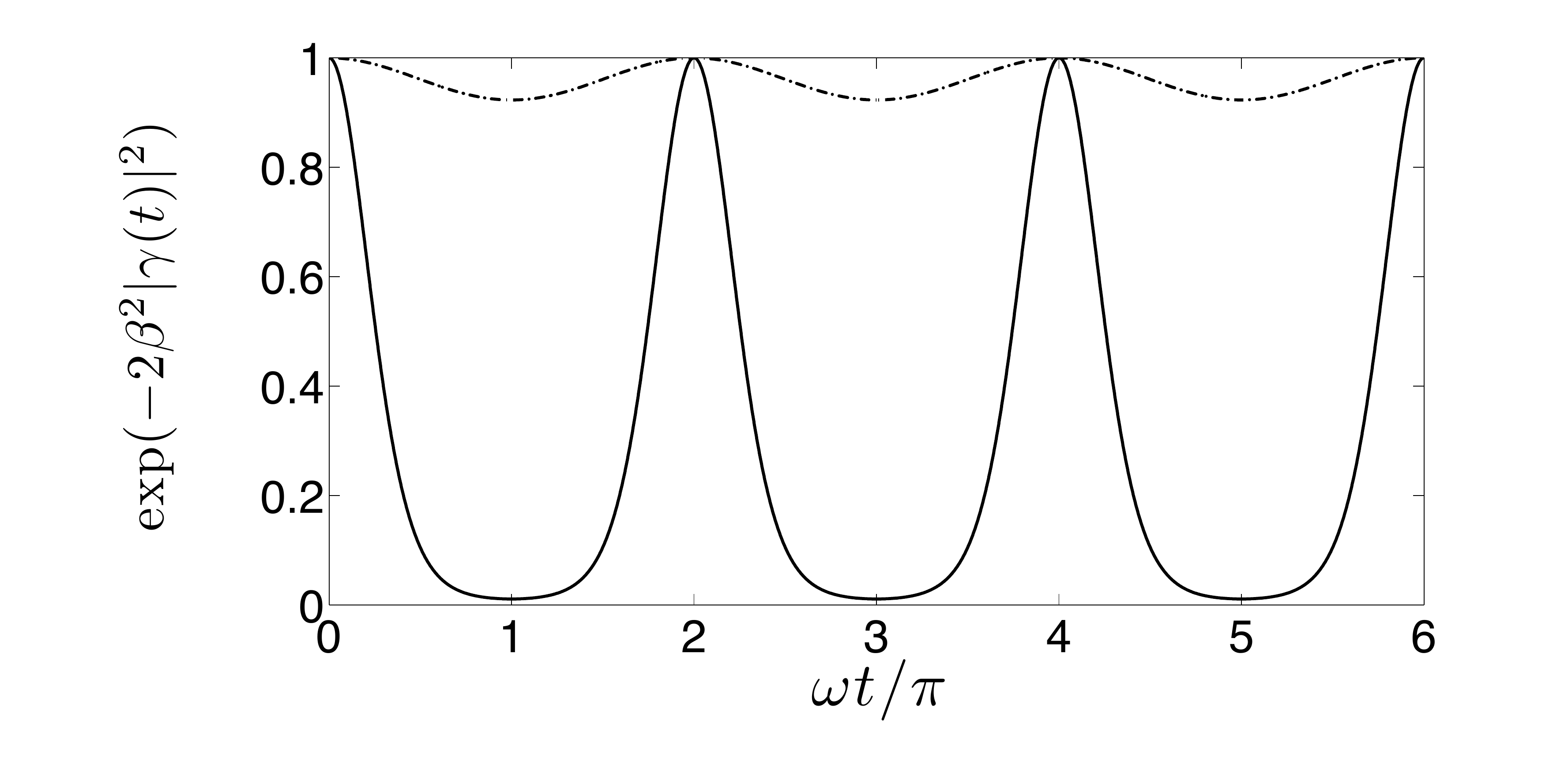}
\caption{\label{gammapic} The time evolution of exp$(-2\beta^{2}|\gamma(t)|^{2})$ for (solid line) $\beta=0.75$ and (dotted-dashed line) $\beta=0.1$.}
\end{figure}
The integral in Eq.(\ref{salam7}) is a two dimensional Fourier transform of $P$. So for each initial  field, one needs to use its  corresponding $P$ representation and calculate its two dimensional Fourier transform and hence $Q(t)$.  In Eq.(\ref{salam7}) all time dependences are captured in terms of $\gamma(t)$ and $\gamma^{*}(t)$. This guarantees that irrespective of the 
initial state the value of Eq.(\ref{salam7}) is a periodic function with the period $T=2\pi/\om$. The periodic dynamics is a manifestation of the 
fact that, in the degenerate regime, the JC Hamiltonian becomes a doubly degenerate harmonic oscillator.  Furthermore, the periodic modulating factor, 
$\text{exp}({-2\beta^{2}|\gamma(t)|^{2}})$,  is present irrespective of the initial field state. In Fig. \ref{gammapic} we have plotted the evolution of this factor for different values of $\beta$. This factor comes from the inner product between two coherent
states in Eq.(\ref{salam6}). These two coherent states came from the evolution of $\ket{\up,\alpha}$ and  $\ket{\dn,\alpha}$. The average complex excitation amplitudes
of these two coherent states are $\alpha$ initially. As time increases these average excitation amplitudes follow two different circles in the complex
plane and if $\beta \gtrsim  0.75$ these
coherent states become effectively orthogonal to each other and their inner product becomes much smaller than 1. At  $t=2\pi/\om$ the average excitation
amplitudes become $\alpha$ again and that is when the modulating factor becomes 1. The periodic damping in the modulating envelope (Fig.\ref{gammapic})  is a manifestation of this effective orthogonality. \\
%
%Entanglement DYNAMICS
%
\section{Entanglement Dynamics}\label{sec4}
\begin{figure}[htpb]
\includegraphics[width=\columnwidth]{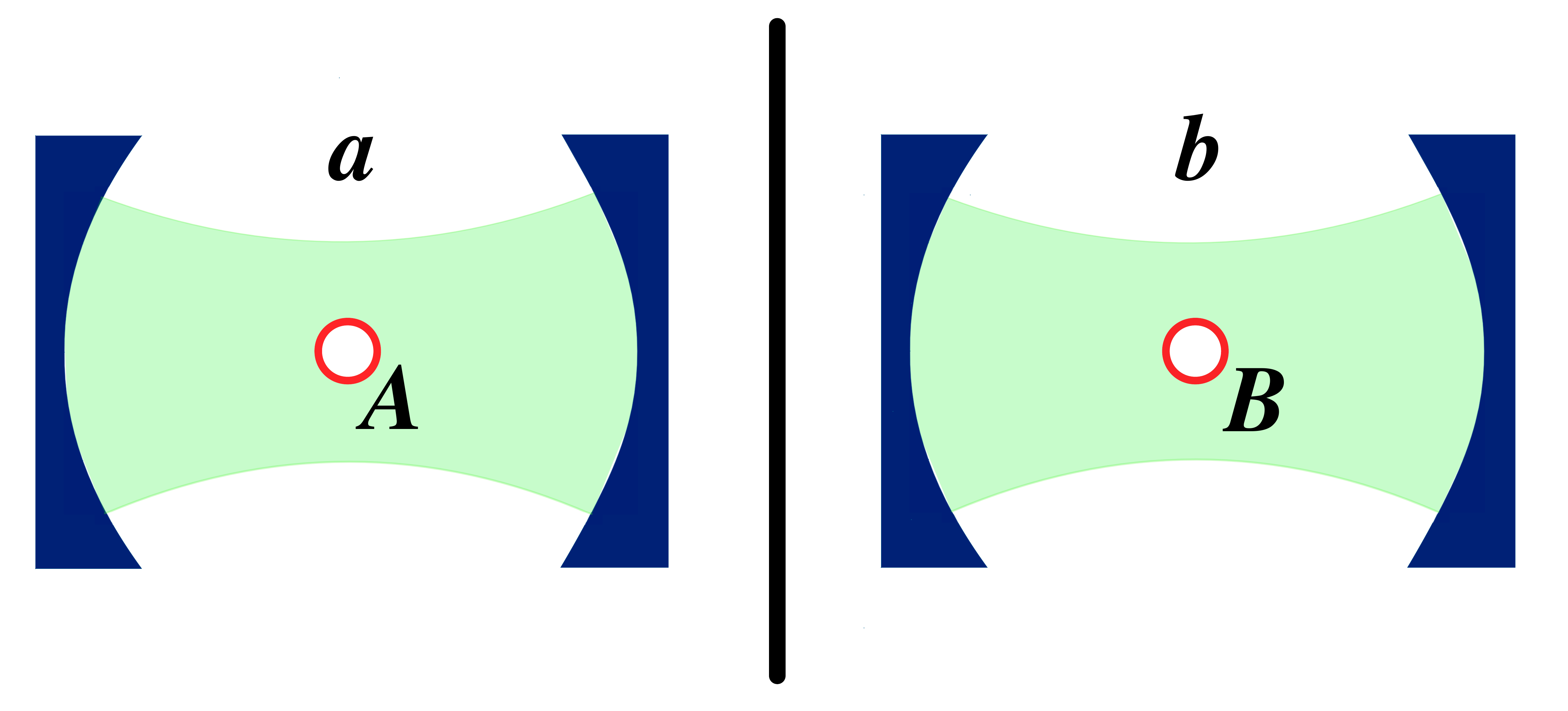}
\caption{A schematic representation of the setup where two remote qubits are interacting with their own environments. There is no interaction between remote parties. Two qubits are assumed to be initially maximally entangled and the fields are in individual separable states.}
\label{schematic}
\end{figure}

%In this section we will study the entanglement dynamics and its transfer in degenerate regime. 
The system we choose to study consists of two non-communicating subsystems, labeled as $Aa$ and $Bb$. Each subsystem itself consists of a qubit, labeled as $A$ and $B$, each interacts with a partner field, labeled as $a$ and $b$. In figure Fig. \ref{schematic} we show a schematic representation of the setup whose entanglement dynamics we seek to analyze. A particular point for attention is whether the initial entanglement shared between two remote systems dies out in a finite period of time, a phenomenon called early stage disentanglement or entanglement sudden death (ESD) \cite{Yu30012009}. 

The Hamiltonian of the system can be written
as the sum of the Hamiltonians of each subsystem. The Hamiltonian of each subsystem is a JC Hamiltonian in the degenerate regime: 
 \begin{align}
H=2\hbar\frac{\la}{\om^{2}}+\sum_{i=1,2}\Bigg(\hbar\omega \dg{a}_i a_i +\hbar g (a_i^{\dagger}+a_i)\sigma_{x}^{(i)}\Bigg).
\end{align}
Two qubits, $A$ and $B$, are assumed to be initially maximally entangled and separable from the fields. After $t=0$ each qubit interacts with its partner field. 

%This setup has been invoked previously to study non-local entanglement dynamics between cavities \cite{yonac1,yonac2007,0953-4075-42-6-065507} in RWA regime. We investigate the entanglement between two qubits after $t=0$ and focus on the effect of different initial fields on this entanglement in the degenerate regime. The fields that are examined are coherent, Fock state or thermal fields. In the RWA regime as qubits interact with their fields some of the information of the qubits gets copied on the fields that leads to a non-local bipartite entanglement between two fields \cite{yonac2007}. In the end of this section it will be shown that in degenerate regime immaterial of how strong the coupling is, no bipartite entanglement develops between two fields.

The initial qubit states under consideration are \begin{align}\ket{\Phi_\pm}=\frac{\ket{e,e}\pm\ket{g,g}}{\sqrt{2}}.\end{align} These are the maximally entangled Bell states that were invoked in previous investigations \cite{yonac1,yonac2007} for a similar scenario in  the RWA regime \cite{yonac1}. Two other Bell states \begin{align}\ket{\Psi_{\pm}}=\frac{\ket{e,g}\pm\ket{g,e}}{\sqrt{2}}\end{align} are not considered separately since in the degenerate regime there is no difference between the entanglement dynamics of $\ket{\Phi_{\pm}}$ and $\ket{\Psi_{\pm}}$. To see the reason for this note that
\begin{align}
\sigma_{x}^{(1)}U\ket{\Phi_{\pm}}\otimes \ket{\Lambda_{a},\Lambda_{b}}=U\ket{\Psi_{\pm}}\otimes \ket{\Lambda_{a},\Lambda_{b}},
\end{align}
where we used the fact that $[\sigma^{(1)},H]=0$ and $\ket{\Lambda_{a},\Lambda_{b}}$ can be any combination of  initial pure field states. Thus, there is a local unitary transformation that brings the state that is produced by the initial state $\ket{\Phi_{\pm}}$  to the state that is the result of the propagation of $\ket{\Psi_{\pm}}$. This
means that these two states have the same entanglement \cite{RevModPhys.81.865}. This result can be readily generalized for all initial states. Throughout this section, where it is needed to quantify entanglement, we take advantage of Wootters concurrence \cite{wootters} as is already used in the RWA regime \cite{yonac1}. 

To find the reduced density matrix of two qubits, $Q^{\pm}(t)$, one can generalize the technique we employed in the previous section. We assume that at $t=0$ the qubits are entangled with each other but are separable from the fields and furthermore, two fields are initially separable too. Thus, the density matrix of the system can be written as
\begin{align}\rho_{ABab}^{\pm}(0)=\ket{\Phi_{\pm}}\bra{\Phi_{\pm}}\otimes F_{a}\otimes F_{b}.\end{align}
If we use the $\ket{\up\up}$,$\ket{\up\dn}$,
$\ket{\dn\up}$ and $\ket{\dn\dn}$ basis, the diagonal terms of $Q^{\pm}(t)$ remain constant. In $Q^{+}(t)$ the only non-vanishing off-diagonal elements are $Q_{\up\up,\dn\dn}^{+}(t)$ and its conjugate. In $Q^{-}(t)$, however, the only non-vanishing off-diagonal elements are 
$Q_{\up\dn,\dn,\up}^{-}(t)$ and its conjugate. Here, for simplicity, we assume the initial fields are identical. If the $P$ representation of the initial fields is $P(\alpha)$, then it can be shown that

\begin{align}\nn
Q_{\up\up,\dn\dn}^{+}(t)&=\frac{e^{-4\beta^{2}|\gamma(t)|^{2}}}{2}\left(\int \text{d}^{2}\alpha P(\alpha)e^{4i\beta Im[ \alpha \gamma^{*}(t)]}\right)^{2},\\
Q_{\up\dn,\dn\up}^{-}(t)&=\frac{e^{-4\beta^{2}|\gamma(t)|^{2}}}{2}~\left|\int \text{d}^{2}\alpha P(\alpha)e^{4i\beta Im[ \alpha \gamma^{*}(t)]}\right|^{2}.
\label{salam8}
\end{align} 
Now for each choice of initial field one can evaluate the integral and through it find $Q(t)$. For the above initial conditions, the concurrence also takes
a very simple form:
\begin{align}C_{AB}^{+}(t)=C_{AB}^{-}(t)=2|Q_{\up\up,\dn\dn}^{+}(t)|=2|Q_{\up\dn,\dn\up}^{-}(t)|.\end{align} The fact that concurrence depends only on the absolute value of  $Q_{\up\up,\dn\dn}^{+}(t)$ guarantees that
the evolution of concurrence is the same for $\ket{\Phi_{\pm}}$ states. For simplicity from now on we drop the subscript $AB$  and superscripts $\pm$ and focus only on the state
$\ket{\Phi_{+}}$.  \\
 \begin{figure*}[tpb]
\includegraphics[width=17cm]{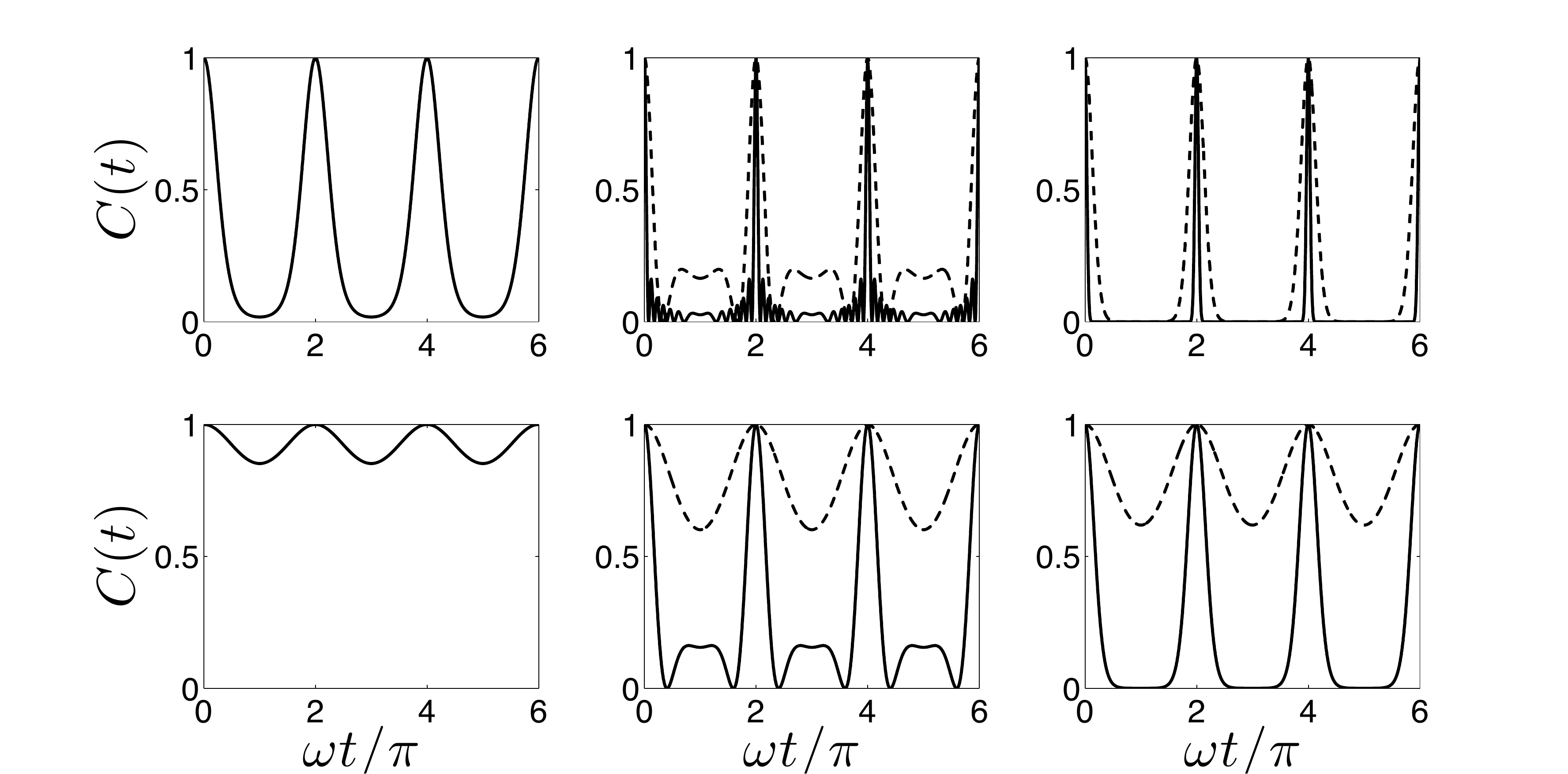}
\caption{\label{mainfig}The evolution of concurrence between two qubits in the degenerate regime for different initial fields and different coupling strength. In the top row we used $\beta=0.5$ and for the bottom row $\beta=0.1$. In the left column the two initial fields are assumed to be in coherent states. In the middle column two initial fields are initially in number states (solid) N=25 and (dashed) N=1. In the right column the initial fields are initially in thermal fields with (solid) $\bar{n}=25$ and (dashed) $\bar{n}=1$ }
\end{figure*}

\textbf{Coherent states}:  We assume two fields are initially in identical coherent states $\ket{\alpha_{0}}$. The initial state of the whole system then reads
$$\ket{\Phi_{+}}\otimes\ket{\alpha_{0}}\otimes\ket{\alpha_{0}}.$$

%In
%\cite{yonac2010} a similar scenario was investigated in the RWA regime with the initial qubits in state $\ket{\Psi_{+}}$. They found the initial entanglement between two fields collapses to zero and stays zero for a finite period of time before it partially revives and this cycle of collapse and revival goes on until the revivals becomes temporally inseparable and dies out. 
In the degenerate regime the off-diagonal term, $Q_{\up\up,\dn\dn}(t)$,  reads
\begin{align}Q_{\up\up,\dn\dn}(t)=\reci{2}e^{8i\beta Im{[\alpha_{0}\gamma^{*}(t)]}}e^{-4\beta^{2}|\gamma(t)|^{2}}.\end{align}
Thus, $\alpha_{0}$ only appears in a phase and this means the average excitation number of a coherent state does not have any effect on the concurrence. Thus, in the degenerate regime, the concurrence between two qubits that are coming in contact with two coherent fields, $C_{coh}(t)$, is the same as two qubits interacting with two initial vacuum fields. In Fig. \ref{mainfig} (left panel), we present the evolution of concurrence between qubits  for a coherent state. As $\beta$ increases, the minimum of the concurrence decreases, but the change of $\alpha_{0}$ does not have any effect on the concurrence. At $\om t=2k\pi$, where $k$ is a natural number, $\gamma(t)=0$ and the initial
entanglement revives completely. This is a result of harmonic oscillation in the dynamics and independent of the initial state.    \\

\textbf{Number states}: Next we focus on the initial fields being number states. Here also for simplicity we assume both initial fields are identical and the initial state is $$\ket{\Phi_{+}}\otimes\ket{N}\otimes\ket{N}.$$

% If the qubits are on resonance with the fields  (${\om_{0}\over\om} =1 $) and RWA is adopted then the concurrence between the two qubits is given by
%\begin{align*}
%C_{N}^{RWA}(t)=\max\{0,\cos^{2}\Omega_{N+1}t\cos^{2}\Omega_{N}t\\-\reci{4}\sin^{2}2\Omega_{N+1}t-\reci{4}\sin^{2}2\Omega_{N}t\}
%\end{align*} 
%where $\Omega_{N}=\la\sqrt{N}$. In Fig.\ref{numberrwa} we plotted  $C_{N}^{RWA}(t)$ for $N=25$. For $N\gg1$ one has $\Omega_{N+1}-\Omega_{N}\simeq\frac{\la}{2\sqrt{N}}\ll1$  and this leads to the beat frequency in Fig. \ref{numberrwa}. For small $N$, $\Omega_{N+1}$ and $\Omega_{N}$ are not close and the entanglement dynamics show aperiodic behavior. For $N=0$ then $\Omega_{0}=0$ and $C_{N}^{RWA}$ becomes a periodic function with the Rabi frequency $\la$ and we recover the result in \cite{yonac1}.

To derive the entanglement dynamics in the degenerate regime one can use the $P$ representation of a number state \cite{scully}:

\begin{align*}P_{N}(\alpha)=\frac{e^{\alpha\alpha^{*}}}{N!}\frac{\partial^{2N}}{\partial \alpha^{N}\partial\alpha^{*N}}\delta^{(2)}(\alpha)\end{align*}
 and evaluate the integral in Eq.(\ref{salam8}). It can be shown that 
\begin{align}C_N(t)=e^{-4\beta^{2}|\gamma(t)|^{2}}[L_{N}(4\beta^{2}|\gamma(t)|^{2})]^{2},\end{align} where $L_{N}$ is a Laguerre polynomial. In 
Fig. \ref{mainfig} (middle column) we plotted $C_{N}(t)$ for different values of $N$. The complete restoration of the entanglement is again a signature of the harmonic oscillatior dynamics. As $\beta$ increases, $4\beta^{2}|\gamma(t)|^{2}$ reaches more zeros of $L_{N}$ in a period of oscillation and thus concurrence vanishes momentarily at more points of time.  $L_{N}$ has only $N$ roots and thus the number of moments at which concurrence vanishes is at most $2N$.
 As $N$ increases, the two qubits  spend less time remaining maximally entangled. Thus, for a thermal field which is a mixture of 
Fock states one would also expect the temporal width of the restoration to decrease as the average excitation number of the two fields increases. \\

% \begin{figure}[htpb] 
%\includegraphics[width=\columnwidth]{deg_number.pdf}
%\caption{The evolution of concurrence between two qubits in degenerate regime. The initial
%fields are identical number states with (solid line)$N=25$ and (dashed line) $N=0$. For both cases $\beta=0.5$.}
%\label{numberdeg}
%\end{figure}
\textbf{Thermal states}:
In this section we assume that two fields are initially in identical thermal states. Thermal environments
are typically associated with the loss of coherence in the system when it comes in contact with the environment. However,  due to the harmonic nature of the degenerate regime we do expect a complete restoration of the concurrence when $\gamma(t)=0$. To study the entanglement dynamics in the degenerate regime we can invoke the $P$ representation of a 
thermal field \cite{scully}:

%In Fig. \ref{thermalrwa}  we plotted $C_{th}^{RWA}(t)$ for different values of average excitation numbers, $\bar{n}=0.02,0.9$ in the RWA regime and on resonance $(\om_{0}=\om)$. As $\bar{n}$ increases the evolution of $C_{th}^{RWA}(t)$ becomes aperiodic. 

% \begin{figure}[htpb]
%\includegraphics[width=7.5cm]{thermal_rwa.pdf}
%\caption{The evolution of concurrence between two qubits in RWA regime and on resonance ($\om_{0}=\om$). The initial
%fields are identical thermal states with (dashed) $\bar{n}=0.02$ and (solid) $\bar{n}=0.9$.}
%\label{thermalrwa}
%\end{figure}

$$P_{th}(\alpha)=\reci{\pi \bar{n}}\text{exp}(-|\alpha|^{2}/\bar{n}).$$
The integral in Eq.(\ref{salam8}) becomes a Gaussian integral. For two initially identical fields with average excitation $\bar{n}$, the concurrence between two qubits in the degenerate regime is given by
\begin{align}
C_{th}(t)=\text{exp}(-4 (1+2\bar{n})\beta^{2}|\gamma(t)|^{2}).\label{salam9}
\end{align}
As argued before, by increasing the average excitation number of thermal fields the temporal width of the restoration period decreases. This can also be understood
in terms of coherent states. To the off-diagonal elements of
the density matrix of two qubits, each coherent state contributes a specific phase $\text{exp}(8i\beta Im{[\alpha\gamma^{*}(t)]})$. As $\bar{n}$ increases the width of the Gaussian in the complex plane increases and more $\alpha$'s contribute significantly to the integral in Eq.(\ref{salam8}). This leads to a faster
collapse of $C_{th}(t)$ as $\gamma(t)$ becomes non-zero. In other words as a Gaussian broadens in the $\alpha$ domain, its Fourier transform becomes narrower. From Eq.(\ref{salam8}) one can conclude that the effect of the average excitation number in the thermal field is the same as to enhance the coupling $\beta$.  In Fig. \ref{mainfig} (right column)  $C_{th}(t)$ is presented for different values of $\bar{n}$. It is interesting that for a weak coupling 
and far from the resonance if the fields are thermally excited enough, i.e. $8\bar{n}\beta^{2}\gtrsim1$, then the fields effect both the local-coherences and entanglement between two qubits considerably in a short time.

\begin{figure}[htpb]
\includegraphics[width=\columnwidth]{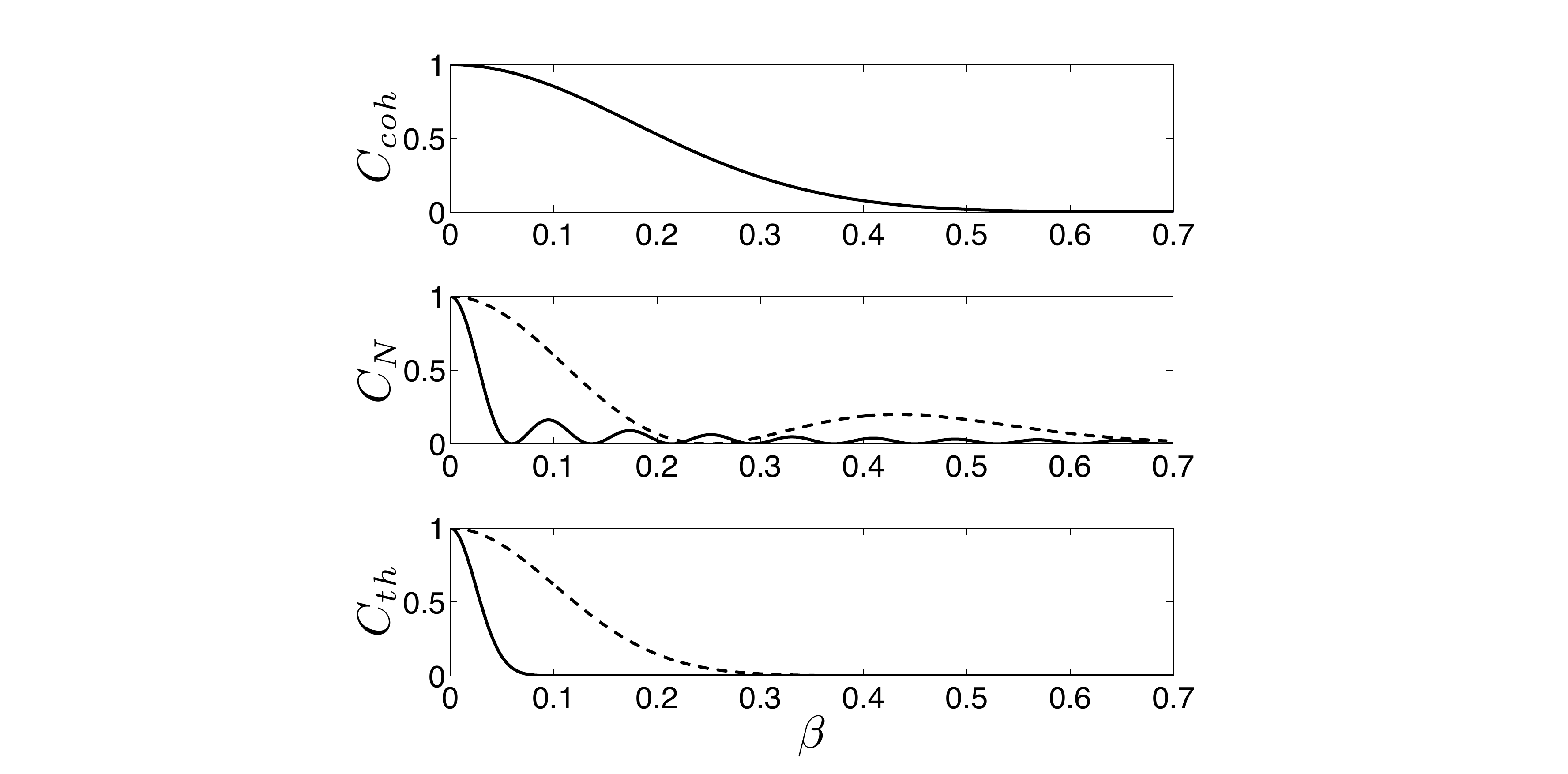}
\caption{ Concurrence between two qubits, evaluated at $t=\pi/\om$, as a function of $\beta$. Top, we presented the concurrence when the initial fields are identical coherent states. Middle row, presented is the concurrence when the initial fields are number states (solid) N=25 and (dashed) N=1. Bottom, we presented the concurrence when the initial fields are identical thermal fields (solid) $\bar{n}=25$ and (dashed) $\bar{n}=1$. }
\label{betadependence}
\end{figure}

For all the initial states we studied so far, the concurrence is an explicit function of $4\beta^{2}|\gamma(t)|^{2}$. This control quantity is a periodic function in time and it also depends explicitly on $\beta$. Our treatment of the degenerate regime does not impose any constraint on the value of $\beta$. Here we fix a time and study the effect of the coupling strength on the concurrence. We study the effect of $\beta$ at time $t=\frac{\pi}{\om}$, when the control quantity, $4\beta^{2}|\gamma(t)|^{2}$, takes its maximum value $16\beta^{2}$.  One can show that at $t=\frac{\pi}{\om}$
\begin{align}\nn
C_{coh}(\pi/\om)&=e^{-16\beta^{2}},\\ \nn
C_{N}(\pi/\om)&=e^{-16\beta^{2}}[L_{N}(16\beta^{2})]^{2},\\
C_{th}(\pi/\om)&=e^{-16\beta^{2}(1+2\bar{n})}.
\end{align}
 In Fig. \ref{betadependence} we plot the concurrence at $t=\frac{\pi}{\om}$ versus $\beta$. As $\beta$ increases the concurrence decreases exponentially. The initial excitation in coherent states, (top panel), does not effect the concurrence. In the middle panel we plot the concurrence for the number state fields. The effect of the number state excitation is captured in the Laguerre polynomial modulation of the exponential decay. As we see in the bottom panel, increasing the average excitation number in thermal fields enhances the exponential decrease of the concurrence. 

 \begin{figure}[tpb]
\includegraphics[width=\columnwidth]{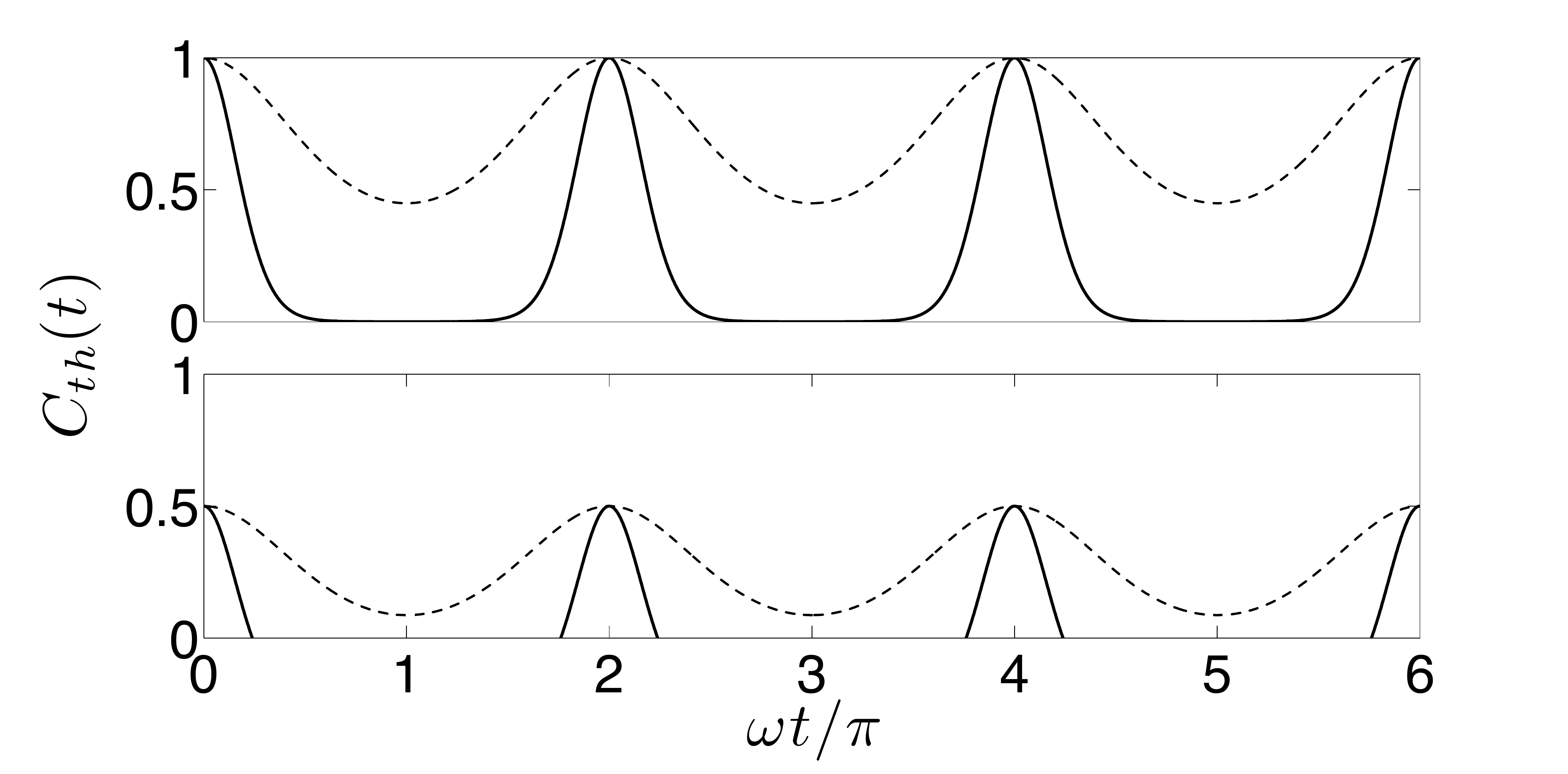}
\caption{ The evolution of concurrence between two qubits in the degenerate regime for $\beta=0.1$. In the top the initial
fields are identical thermal fields with (solid) $\bar{n}=25$ and (dashed) $\bar{n}=2$. In the bottom we presented $C_{th}(t)$ for the 
initial thermal fields and qubits are initially in the state described in Eq.(\ref{salam10}) in degenerate regime.
 (solid) $\bar{n}=25$ and (dashed) $\bar{n}=2$. }
\label{thermaldeg}
\end{figure}

Next we focus on the phenomenon of ESD \cite{Yu30012009}. From Eq.(\ref{salam9}) one sees that $C_{th}(t)>0$, so irrespective of how strong the coupling is, ESD does not happen in the degenerate regime. Increasing the 
coupling strength only decreases the minimum of the entanglement. It shall be pointed out that this is an artifact of the initial qubit state that was chosen and is not a generic property of the degenerate regime. The reason that this initial state does not show ESD is that in the basis $\ket{\up\up}$,$\ket{\up\dn}$, $\ket{\dn\up}$ and $\ket{\dn\dn}$, two of the diagonal terms of the density matrix $Q(t)$ are zero. Thus to have ESD the off-diagonal element, $Q_{\up\up,\dn\dn}(t)$, should vanish for a finite time interval, but this element does not vanish. 

In Fig. \ref{thermaldeg}, along with $C_{th}(t)$ for $\ket{\Phi_{+}}$, we also plot $C_{th}(t)$ when the qubits share the initial state 
\begin{align}
Q(0)=\frac{3}{4}\ket{\Phi_{+}}\bra{\Phi_{+}}+\reci{8}\ket{\up\dn}\bra{\up\dn}+\reci{8}\ket{\dn\up}\bra{\dn\up}.\label{salam10}
\end{align}
For this initial state, if $\beta$ is big enough, two qubits become disentangled for a finite period of time and thus ESD happens.

So far we studied the evolution of the bipartite entanglement between two qubits and focused on the effect of the local environments on $C(t)$.
The question that arises is how does the this entanglement get transferred between the different parties involved? In the RWA regime, the question is 
answered for the case when the initial fields are in the vacuum \cite{yonac2007}. They showed that after $t=0$ the interaction between each qubit 
and the corresponding field leads to the development of non-local bipartite entanglement  between two non-local fields. In what follows we show
that in the degenerate regime, irrespective to how strong the coupling is, and what the initial fields are, the two initially separable non-local fields do not develop bipartite entanglement. 

To this end, assume the initial fields to be two vacuum fields so that the initial state of the system reads
\begin{align}\ket{\Phi_{+}}\otimes\ket{0,0}=\reci{\sqrt{2}}\Big(\ket{\up\up}+\ket{\dn\dn}\Big) \otimes \ket{0,0}.\end{align}
After $t=0$, each qubit interacts with the corresponding field. Therefore at $t>0$ the state of the system is given by
\begin{align}
U\ket{\Phi_{+}}\otimes\ket{0,0}= \frac{\ket{\up\up,\beta(t),\beta(t)}+\ket{\dn\dn,-\beta(t),-\beta(t)}}{\sqrt{2}},
\end{align}  
where $\ket{\pm\beta(t)}$ are coherent states and $\beta(t)=\beta(e^{-i\om t}-1)=\beta\gamma^{*}(t)$. For the above state if two qubits are traced out, the reduced density matrix of the two fields is separable.  In other words at any moment an observer can measure the state of the qubits in the basis $\ket{\up\up}$, $\ket{\up\dn}$, $\ket{\dn\up}$ and $\ket{\dn\dn}$ and by knowing the result one can also tell the state of each field. One can readily generalize the above result to any initial state for which the fields are separable from each other and from the qubits.

The question that remains is the destination of initial entanglement. To where is it transferred? For the initial condition we studied above, if the coupling 
is strong enough such that $\ipr{\beta(t)}{-\beta(t)}\approx 0$, then the initial bipartite entanglement between two qubits becomes a pure 4-partite
entanglement between all the parties involved and there remains no bipartite entanglement  in the system. 
 \section{conclusion}
  In this report we studied the excitation exchange and entanglement dynamics in the Jaynes-Cummings model far from the RWA regime.
 In the degenerate regime, the dynamics can be understood as displaced harmonic oscillations of the field around a center that depends on the qubit state. This leads to complete restoration of coherences irrespective of the initial state.

  We also invoked a previously studied model \cite{yonac2007} and studied the entanglement dynamics for two remote qubits that are interacting with two local environments in the degenerate regime. We assumed that initially the qubits are separable from the environments and of all initial qubit states we  chose to focus on the Bell states. It was shown that $\ket{\Psi_{\pm}}$ has the same entanglement dynamics as $\ket{\Phi_{\pm}}$. Different choices of single mode environments were examined. We showed that the effect of all coherent states on the concurrence is the same as the effect of the vacuum state and initial excitation in a coherent state does not have any effect on the bipartite entanglement between two qubits. In cases of number state and thermal fields the initial excitation of the fields does effect the evolution
  of concurrence between two qubits. In the case of thermal fields, the effect can be captured as an enhancement of the coupling between each qubit and the corresponding field. The fact that a highly excited thermal field can, in a short time, affect the local and non-local coherences of a degenerate off-resonance qubit that is weakly coupled to it and a highly excited coherent state can not, is of importance.
  
  In another  sharp contrast to the previously studied scenario \cite{yonac2007},  it was shown that no bipartite entanglement can be 
  induced between two remote fields using the initial entanglement between two qubits in degenerate regime. This raises a question about the quasi-degenerate regime that remains to be considered for future investigation. The question is, for the regime where the degeneracy of the qubits is broken with a small splitting $\omega_{o}\ll\om,\la$, can the initial entanglement between two qubits induce  a non-zero bipartite entanglement between two fields. If yes, then is there a limit on the amount of this induced entanglement or not, and in what time scale does the entanglement get transferred? Finally, for the initial state that we examined, all the initial entanglement transformed to a 4-partite entanglement between all parties involved.  The presented scenario can also be thought of as a scenario to produce pure 4-partite entanglement which is potentially useful in the studies of multi-partite entanglement.
 
% We also prompted a previously studied model to study the entanglement dynamics in degenerate regime and compare it with the RWA regime. It was shown that irrespective of how strong the coupling is in the degenerate regime the atoms can not induce bipartite entanglement between remote fields. This result invokes a question that should be investigated in the future investigation. The question is if the degeneracy breaks with a very small amount, $\om_{0}\ll\om$, then two fields can perhaps develop some bipartite entanglement during dynamics. The question to answer is then what is the maximum of that entanglement and is it independent of the strength of the coupling. We also studied the effect of the initial excitations in the fields on the atom-atom entanglement. We showed for the states we studied the coherent states does not have any effect on the concurrence. A number state and a thermal state will have some effect on the bipartite atom-atom entanglement and for a thermal state the effect can be thought of as an enhancement of the coupling.
 
\section{acknowledgement}
We acknowledge partial financial support from ARO W911NF-09-1-0385 and NSF PHY-0601804. \\
\section{Appendix A}
In this section we prove the equation \ref{salam5}. We are interested in finding the evolution of $\ket{\up,\alpha}$. Note that
\begin{align*}
&U\ket{\up,\alpha}=\ket{\up}\otimes D(\beta)^{\dagger}e^{-i\om t \dg{a}a}D(\beta)\ket{\alpha}\\
&=\ket{\up}\otimes D(\beta)^{\dagger}e^{-i\om t \dg{a}a}\ket{(\alpha+\beta)}e^{i Im{(\beta\alpha^{*})}}\\
&=\ket{\up}\otimes D(\beta)^{\dagger}\ket{(\alpha+\beta)e^{-i\om t}}e^{i Im{(\beta\alpha^{*})}}\\
&=\ket{\up}\otimes \ket{(\alpha+\beta)e^{-i\om t}-\beta}e^{-i\beta^{2} \sin\om t} e^{-i\beta Im(\alpha^{*}(e^{i\om t}-1))}\\
&=\ket{\up,(\alpha+\beta)e^{-i\om t}-\beta}e^{-i\beta^{2} \sin\om t} e^{\nim\beta\alpha\gamma^{*}(t)}e^{-\nim \beta \alpha^{*} \gamma(t)}\\
&=\ket{\up,(\alpha+\beta)e^{-i\om t}-\beta}e^{-i\beta^{2} \sin\om t}e^{i\beta Im[\alpha\gamma^{*}(t)]}.
\end{align*}
In deriving the above result it is assumed that $\beta$ is a real number. The evolution of the state $\ket{\dn,\alpha}$ can be 
worked out in a similar way.

\bibliographystyle{apsrev4-1}
\bibliography{mybib}

\end{document}